\documentstyle[12pt,psfig]{article}
\newcommand{\msun}{M_{\odot}}
\title{\bf Interferometry and the study of binaries}
\author{Frank Verbunt\\
\vspace{1cm}\\
\normalsize Sterrenkundig Instituut, Postbox 80\,000,
3508 TA Utrecht, the Netherlands}
\date{\today}
\begin{document}
\maketitle
\pagestyle{empty}
%
% WE REDEFINE THE plain LaTeX PAGESTYLE !!! 
% THIS PAGESTYLE WILL BE USED FOR THE FIRST PAGE ONLY !
%
\def\bull{\vrule height .9ex width .8ex depth -.1ex}
\makeatletter
\def\ps@plain{\let\@mkboth\gobbletwo
\def\@oddhead{}\def\@oddfoot{\hfil\tiny\bull\quad
``Science Case for Next Generation Optical/Infrared Interferometric Facilities 
(the post VLTI era)'';
37$^{\mbox{\rm th}}$ Li\`ege\ Int.\ Astroph.\ Coll., 2004\quad\bull}%
\def\@evenhead{}\let\@evenfoot\@oddfoot}
\makeatother
%
% AND DEFINE OUR MACROS FOR THE REFERENCE LIST
% I.E \beginrefer \refer and \endrefer
%
\def\beginrefer{\section*{References}%
\begin{quotation}\mbox{}\par}
\def\refer#1\par{{\setlength{\parindent}{-\leftmargin}\indent#1\par}}
\def\endrefer{\end{quotation}}
%
% BEGIN THE ABSTRACT CHAPTER WITH \noindent\small, ENCLOSE IT IN A GROUP
% AND BOLDFACE THE TITLE.
%
{\noindent\small{\bf Abstract:} 

To determine the parameters (masses, orbital period) of a binary,
one requires among others the inclination, which is best determined from a
visual orbit. The next generation of interferometers can provide
visual orbits for a large number of binaries. We then can investigate
with what parameters binaries are born, and how these parameters
evolve, and thus understand the galactic binary population. Results
can be obtained quickly by carefully selecting binaries for the study
of evolutionary processes. Full knowledge of properties of binaries at
birth requires a large, dedicated programme, which is best limited to
spectroscopic binaries with primary masses $>0.8M_\odot$ and orbital periods
$<50$\,yr.}
%
% NOW COMES THE MAIN BODY OF THE ARTICLE
%
\section{Introduction}

\begin{figure}
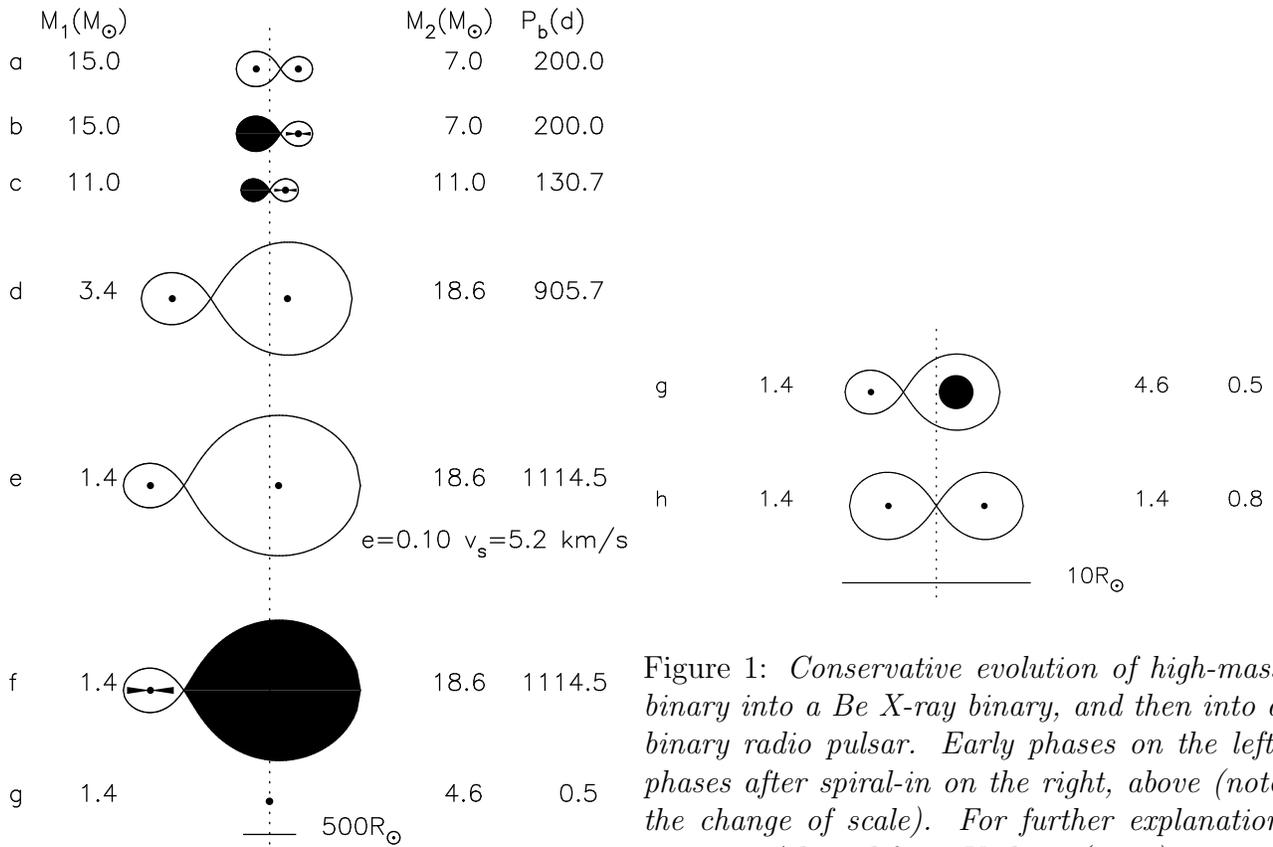

\centerline{
\parbox[b]{8.5cm}{\psfig{figure=verbfaa.ps,width=8.5cm,clip=t}
}
\parbox[b]{8.5cm}{
\psfig{figure=verbfab.ps,width=8.5cm,clip=t}
\caption[o]{\it 
Conservative evolution of high-mass binary into a Be X-ray
binary, and then into a binary radio pulsar. Early phases on the left,
phases after spiral-in on the right, above (note the change of scale).
For further explanation see text. Adapted from Verbunt (1993).
     \label{verbfa}}}
}
\end{figure}

The study of binaries has a number of goals. To introduce these, let
us look at the evolution of the binary shown in Figure\,\ref{verbfa}.
The initial binary consists of stars with masses of 15 and 7 $\msun$,
in an orbit of 200\,d. The more massive star evolves fastest, ascends
the giant branch and transfers its envelope to its companion
(b-d). During this process, the size of the orbit changes due to
conservation of angular momentum. The denuded core of the initially
most massive star continues to evolve and via a supernova event forms
a neutron star (e). The companion to the neutron star through
accretion has gained mass and angular momentum, and has thereby turned
into a Be star. When this star evolves in turn, it engulfs the neutron
star (f-g). The envelope is expelled as the neutron star moves in. At
the end the neutron star forms a close binary with the helium
core. This core in turn becomes a neutron star via a supernova event,
and thus a binary of two neutron stars may be formed (h).

In general it is thought that the full evolution of any binary is set
by the initial conditions, in particular the two masses and the
orbital period.  If the supernova event imparts an appreciable
velocity to the neutron star in an arbitrary direction, an element of
uncertainty is introduced, which means that the evolution following a
supernova event can only be described statistically. Other factors of
interest, but generally less dramatic in importance, are the initial
eccentricity of the orbit, the chemical composition of the stars, and
their initial rotation.  If one would know the distribution of all
these properties for newly formed binaries, one could synthesize a
population of binaries in our galaxy as follows.  In step one, one
chooses randomly from the initial distributions a realization of the
binary, i.e.\ in particular a mass for the more massive star, a mass
for the less massive star, and the orbital period.  One then computes
the evolution of this binary and keeps track of all its different
stages. This procedure is repeated until a sufficiently large number
of binary evolutions is available. The sum of all this provides the
current population of binaries in our galaxy, for a stationary birth
rate. If one is interested in relatively old binaries, one must weigh
the numbers of binaries of different age using the history of the star
formation rate.

%\nocite{ver93}

From this we can derive three important questions that the
study of binaries tries to answer.
\begin{itemize}
\item what are the distributions of the parameters of binaries at birth,
in particular of the primary mass $M_1$, secondary mass $M_2$ and orbital
period $P_b$? This question relates closely to the formation of stars.
\item how do binaries evolve? This question includes the evolution
of single stars, as well as processes typical for binaries, such as
mass transfer and tidal interaction. It relates to the evolution of
the radius of a star with time, to the occurrence of special processes
(type Ia supernovae, gamma-ray bursts), and to the enrichment and
energetics of the interstellar medium.
\item what do we learn from individually interesting systems? As an
example, accurate measurements of masses of neutron stars constrain
the equation of state at nuclear densities (which determines
the maximum possible mass for a neutron star).
\end{itemize}

Before discussing these topics, we first explain how interferometry
helps to obtain information about a binary.

\section{How to study a binary?}

\begin{figure}
\centerline{\psfig{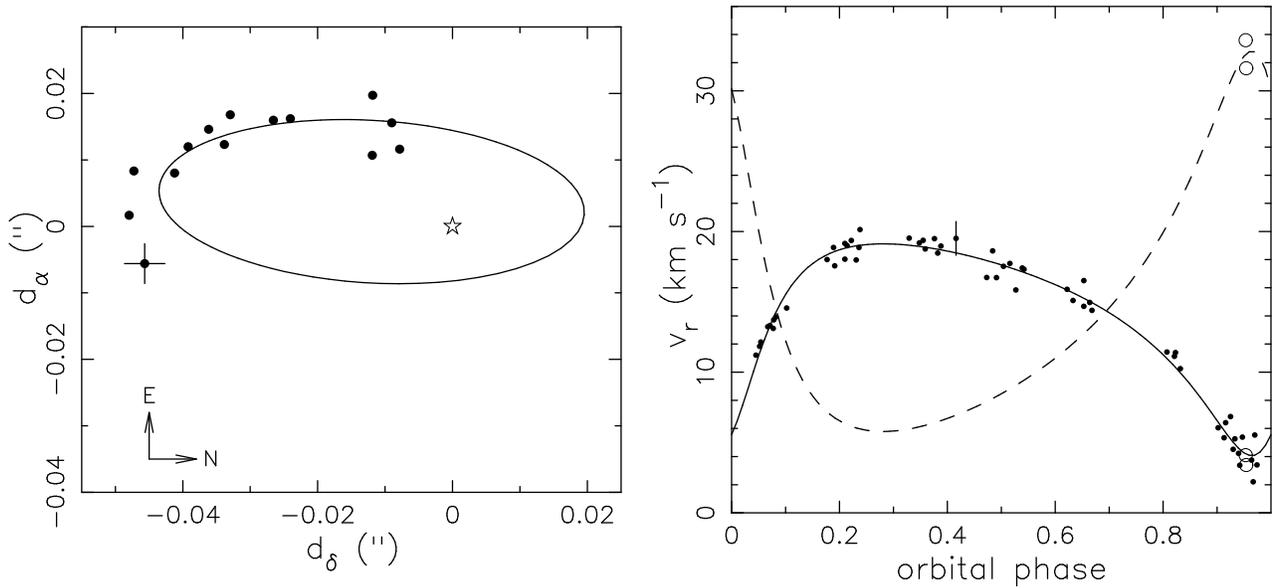}}

\caption[o]{\it Visual orbit (left) and radial velocities (right) 
for a naked T Tauri star (045251+3016) in the Taurus Auriga star forming
region. Note that even with a rough visual orbit, the inclination
is accurately determined, at 113.8$^\circ\pm$3.4$^\circ$. Also note that
once the radial velocity curve of one star is well determined, the
other one can be fixed with one or two accurate measurements (indicated
$\circ$).
After Steffen et al. (2001).
     \label{verbfb}}
\end{figure}

The orbital period $P_b$ and the semi-amplitude of the radial velocity $K_1$
of star 1 in a binary provide an equation, called the mass function,
for three unknowns: the masses $M_1$ and $M_2$ and the inclination $i$ of the
binary with respect to the line of sight ($i=90^\circ$ for edge-on):
\begin{equation}
{P_b{K_1}^3\over2\pi G}\left(1-e^2\right)^{3/2}
 = {(M_2\sin i)^3\over(M_1+M_2)^2}
\end{equation}
Here $G$ is the gravitational constant, and $e$ the orbital eccentricity,
which is determined from the radial velocity curve (sinusoidal for $e=0$).
We thus require two more equations for the three unknowns.
In a double-lined binary, one such equation is the mass function
for star 2. We see that by dividing the two mass functions, we
obtain the mass ratio $q\equiv M_2/M_1= K_1/K_2$. 
This leaves $i$ as an unknown.
There are various methods to determine $i$, but in general the
most accurate method is to measure the visual relative orbit 
(Fig.\,\ref{verbfb}).
As a bonus, this measurement also provides us with the distance
to the binary (from the comparison of astrometric velocities
[$''/s$] and velocities from the spectra [km/s]). The distance is
required if we wish to determine the radii of the stars.
In a single-lined binary, all parameters ($M_1$, $M_2$, $i$)
can be solved if we can measure the orbits of the two stars separately.
For a more exhaustive list of possible ways to study a binary see
Quirrenbach (2001). 
%\nocite{qui01}

For the radii of the stars we can use the methods used for single
stars, provided we can separate their images, c.q.\ determine
the flux of each star separately.

%\nocite{sml+01}

\begin{figure}
\centerline{\psfig{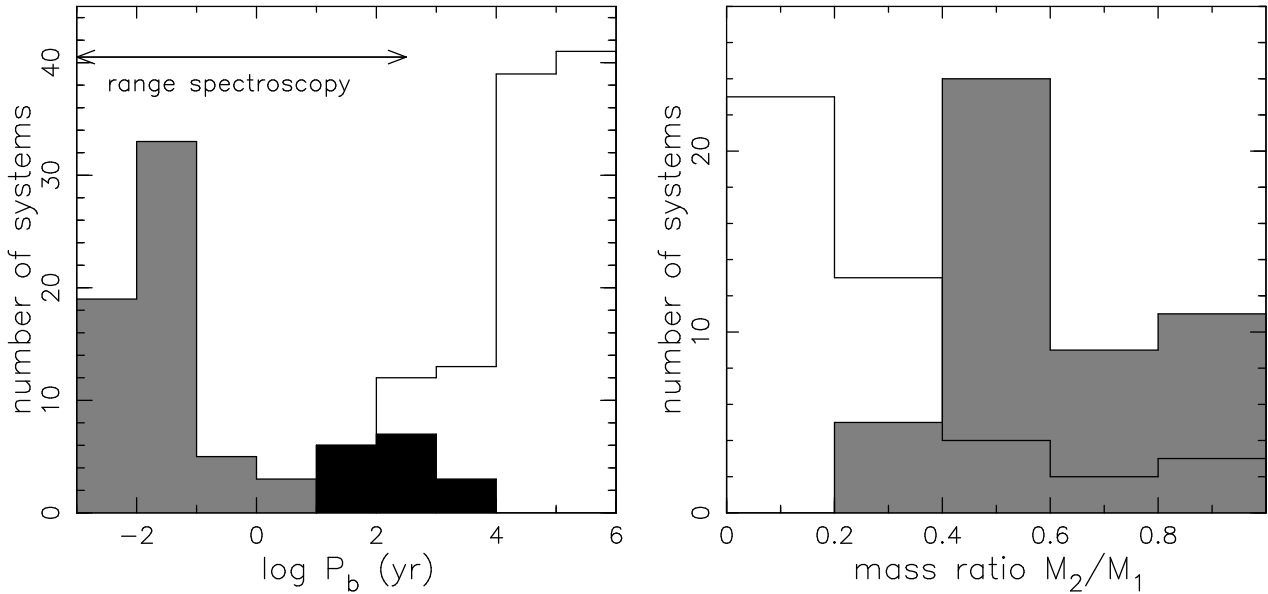}}

\caption[o]{\it Distribution of orbital periods (left) and mass ratios
(right) of O stars as observed (i.e.\ not corrected for selection effects). 
Spectroscopic binaries are indicated with the gray
histogram, visual binaries white, speckle binaries black. Most orbital
periods in excess of $\sim50$\,yr are estimated, from the (projected)
distance between the stars. After Mason et al.\ (1998).
     \label{verbfc}}
\end{figure}

Radial velocities are high in close binaries, separate images of the
binary stars are easiest made of wide binaries (Fig.\,\ref{verbfc}). 
Due to improved
accuracy in the measurements of radial velocities on one hand, and to
higher spatial resolution in imaging, the number of spectroscopic
binaries for which visual orbits can be determined is increasing. It
is here that interferometry helps. Interferometry may produce a direct
image of the binary. A nice example is the bispectrum speckle
interferometry with the SAO 6\,m telescope of OB stars in the Orion
nebula cluster (Schertl et al.\ 2003 and references therein). By doing
this over a long period of time, one can produce the relative visual orbit.
An example from the HST Fine Guidance Sensor is the orbit of a naked
T Tauri star in the star-forming regions of Taurus-Auriga (Steffen et 
al.\  2001, see Fig.\,\ref{verbfb}). 
It is important to note, however, that actual imaging
is not necessary: one can also fit the observed visibilities directly.
A good example is the analysis of 64\,Psc with the Palomar Testbed
Interferometer (Boden et al.\ 1999). This method has the advantage
that shorter observations can be used.
%\nocite{sbpw03}\nocite{sml+01}\nocite{blc+99}

%\nocite{mgh+98}

The examples given above are arbitrarily chosen: an extensive catalogue
of visual orbits is maintained by Hartkopf \&\ Mason at
{\tt http://ad.usno.navy.mil/wds/orb6.html}; see also Hartkopf et
al.\ (2001).
%\nocite{hmw01}

\section{The initial parameters of newly born binaries}

\begin{figure}
\centerline{\psfig{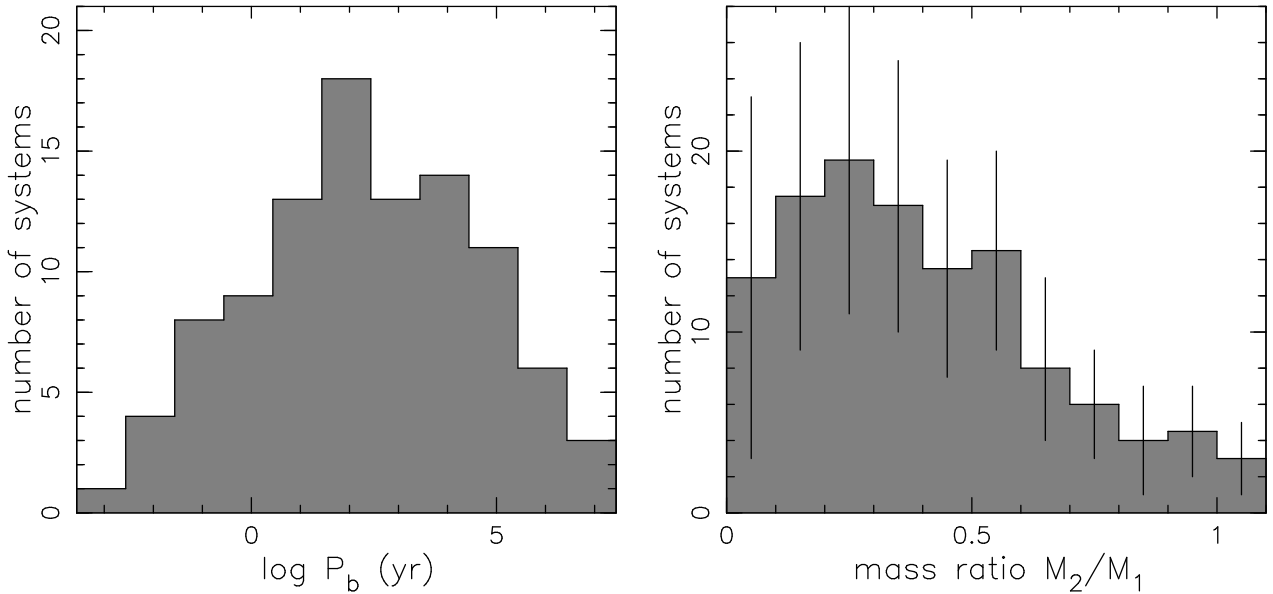}}

\caption[o]{\it Distribution of orbital periods (left) and mass ratios
(right) of nearby G dwarfs. Numbers are corrected for selection effects;
their uncertainties are roughly Poissonian in the period bins, but bigger
-- as indicated in the plot -- in the mass-ratio bins.
After Duquennoy \& Mayor (1991).
     \label{verbfd}}
\end{figure}

Duquennoy \&\ Mayor (1991) published an extensive study of the
initial parameters of G stars, derived from observations of 164
main-sequence F7-G9  stars. They find for example that the period distribution
is given by a log-normal distribution. Due to relatively large
errors, the distribution of mass ratios is less constrained 
(Fig.\,\ref{verbfd}). One would like to obtain distributions with
equal accuracy for the other spectral types.
%\nocite{dm91}

Mathematically the distribution of initial masses $M_1$ and $M_2$ and
the orbital periods $P_b$ can be written as a probability function
$\Pi(M_1,M_2,P_b)$. It is tempting to assume that this combined
probability function can be separated into separate probabilities
for the masses and period, i.e.\ $\Pi(M_1,M_2,P_b)=\Pi(M_1)\Pi(M_2)\Pi(P_b)$,
or alternatively for the mass ratio $q\equiv M_2/M_1$ that
$\Pi(M_1,q,P_b)=\Pi(M_1)\Pi(q)\Pi(P_b)$. 
From the limited observations that we have we already know that 
this is not the case:
$$ \Pi(M_1,q,P_b)\neq\Pi(M_1)\Pi(q)\Pi(P_b); \qquad q\equiv M_2/M_1$$
Compare for example the distributions for O stars 
(Fig.\,\ref{verbfc}) and for G stars (Fig.\,\ref{verbfd}).
The period distribution of G stars peaks at $\sim 100$\,yr; the mass-ratio
distribution at low values. These distributions are not compatible
with the distribution for spectroscopic O-star binaries, even if we take into
account the selection effects which prevent us from measuring binaries
with extreme mass ratios ($M_2<0.2M_1$, say)
spectroscopically (there aren't enough O stars left to change the
distributions much).
On the other hand, O stars in wide binaries more often have
low-mass companions. Thus $\Pi(q)$ may depend both on the primary
mass $M_1$ and on the orbital period $P_b$.

The consequence of this is that one will need to study many binaries
before the general distribution of initial parameters can be
determined with some accuracy.
However, for the study of binary evolution one may concentrate on
binaries with primaries that evolve within the Hubble time, i.e.\ 
with masses $M_1>0.8M_\odot$ (spectral type G\,V or earlier), and
in which the stars influence one another, i.e.\ with binary periods
$P_b<50$\,yr. The other binaries either don't evolve at all, or
if they do, the two stars evolve essentially independently from
one another.

\begin{figure}
\centerline{
\parbox[b]{8.5cm}{\psfig{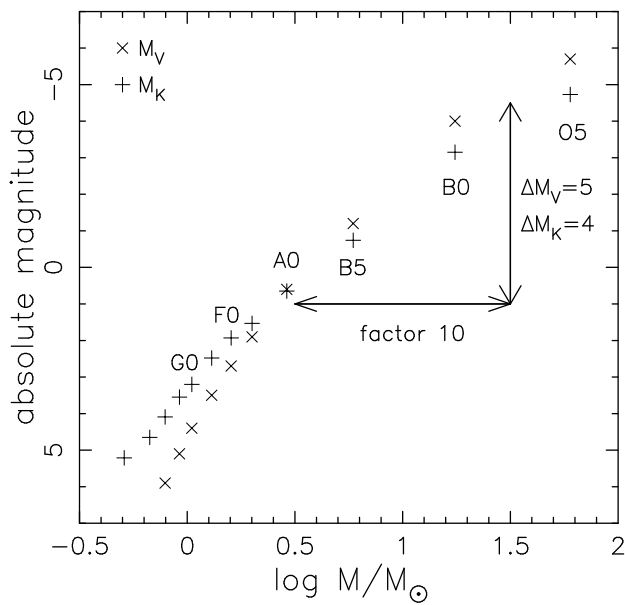}
}
\parbox[b]{8.5cm}{\caption[o]{\it 
Absolute visual and K magnitudes for main-sequence stars. To detect
a companion at one tenth of the mass of the primary, one must bridge
a magnitude difference of $M_V\simeq 5-6$ or $M_K\simeq 4-5$. (Data from
Table 3.13 in Binney \& Merrifield 1998.)
     \label{verbfe}}}
}
\end{figure}
%\nocite{bm98}

With Kepler's law and the definition of the parsec, we may write
the semimajor axis $a$ of a binary, expressed in milliarcseconds,
in terms of the total mass $M\equiv M_1+M_2$, 
orbital period $P$ and distance $d$ as
\begin{equation}
{a\over{\mathrm mas}} = \left({M\over M_\odot}\right)^{1/3}
 \left({P\over 1{\rm yr}}\right)^{2/3} 
 \left({d\over 1{\mathrm kpc}}\right)^{-1}
\label{veq:a}\end{equation}
$a$ changes only slowly with mass -- a range of a factor 10 in $M$
corresponds to a range of 2 in $a$.
{\em I will give $a$ according to this equation as the required resolution},
but the reader should note a) that to resolve the orbit projected on
the sky one would typically need to measure separations down to $\sim 0.2a$,
say; and b) that the visibility doesn't have to be measured down to zero
to measure a separation (I thank F.\ Delplancke for reminding me of this).

To obtain a sufficiently large sample of low-mass stars, of spectral types 
AFGKM, it is enough to reach relatively nearby stellar clusters, including
star formation regions, at a distance of $\sim 130$\,pc, say. With
$M\simeq M_\odot$ we see from Eq.\,\ref{veq:a} that an orbit of 100\,d 
is resolved with 3\,mas at 130\,pc; or with 1\,mas at 400\,pc. 
A G0\,V star has $K\simeq11$ at 400\,pc; an A0\,V star $K\simeq8.5$.
To obtain a sufficiently large sample of OB stars, one needs to study
clusters out to about a kpc; this includes the clusters of the Gould Belt.
At 1\,kpc and for $M=20M_\odot$, 3\,mas resolves an orbit of 1\,yr; and 1\,mas
an orbit of 0.2\,yr.

\begin{figure}
\centerline{\psfig{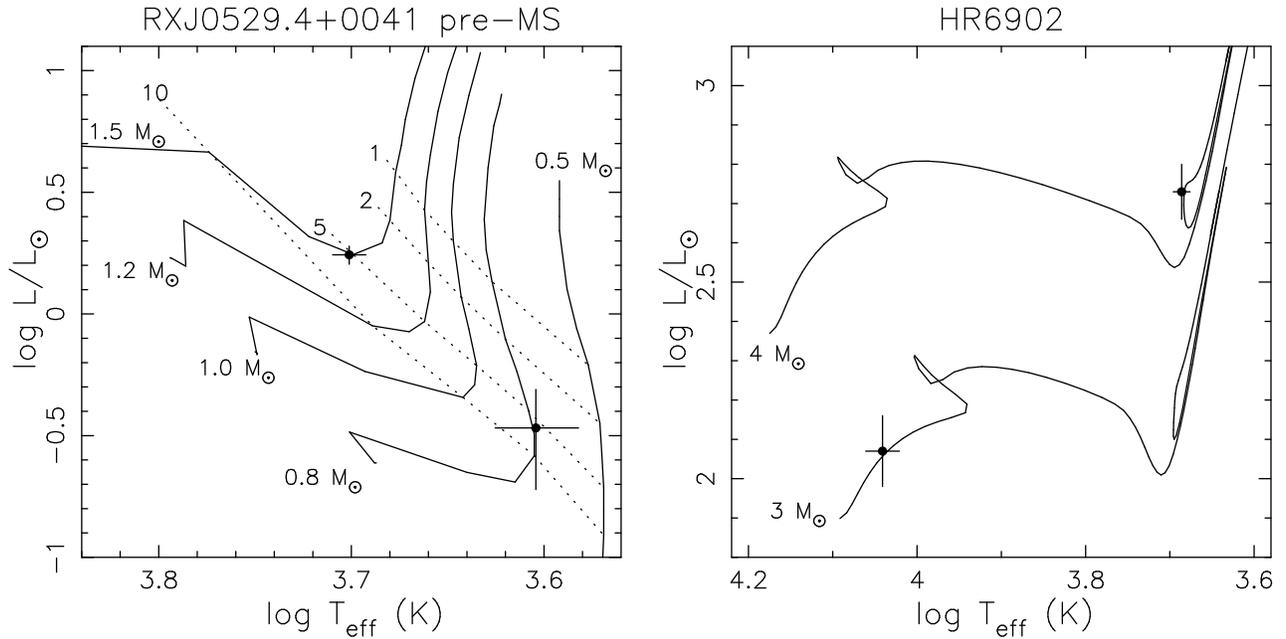}}

\caption[o]{\it Comparison of observed with theoretical location in
Hertzsprung-Russell diagram for a pre-main-sequence binary (left,
after Covino et al.\ 2000, tracks from d'Antona \&\ Mazzitelli 1994), 
and for a normal binary (right, after Schr\"oder et al. 1997).
The dashed lines in the left figure are isochrones, with ages indicated
in Myr.
     \label{verbff}}
\end{figure}

In Fig.\,\ref{verbfe} we plot the visual and K-band magnitudes of
main-sequence stars; from the figure we see that detection of a
companion with one-tenth of the mass of the primary requires a magnitude
contrast of 5-6 in $V$ and slightly less in $K$. A large magnitude
difference makes it more difficult to obtain radial velocities of both
stars. It is therefore useful to note that low-mass pre-main-sequence
stars are relatively bright, and easier detected. As an example, we
see in Fig.\,\ref{verbff} that on the main-sequence a 1.5\,$M_\odot$
star is a factor 25 more luminous than an 0.8\,$M_\odot$ star; but only
a factor 2 when both stars are 1\,Myr old. To study binaries with
extreme mass ratios one may therefore wish to study star-forming regions.

%\nocite{ccf+00}\nocite{dam94}\nocite{spe97}

Improving the resolution of the interferometers by a factor three with
respect to the current resolution increases the number of available orbits
by an order of magnitude for O stars (which are in the galactic plane
and therefore have a two-dimensional spatial distribution), and more
for the less massive stars (which form a thicker disk).

\section{Tests on stellar evolution}

Once we have the parameters of both stars in a binary, we can apply tests
of stellar evolution. The more direct tests are done in a binary where
the stars have not yet influenced one another, and have evolved
essentially independently. We discuss two examples.

The analysis of radial velocities and the visual orbit of the 
pre-main-sequence binary \\
RXJ\,0529.4+0041 (the RX indicates that
the binary is a Rosat X-ray source) give masses of $1.25M_\odot$ and
$0.91M_\odot$, with errors of $0.05M_\odot$. We see 
in Fig.\,\ref{verbff} that the mass derived
for the primary from evolutionary tracks is marginally higher; and that
the positions of the stars in the Hertzsprung-Russell diagram are
compatible with both stars being about 5\,Myr old (Covino et al.\ 2000).
%\nocite{ccf+00}

The stars in HR\,6902 have masses $(3.86\pm0.15)M_\odot$ and
$(2.95\pm0.09)M_\odot$, respectively. The less massive star is still
on the main-sequence, the more massive star has already evolved into a
late-type giant (Fig.\,\ref{verbff}). The locations of the
stars in the Hertzsprung-Russell diagram is compatible with the
evolutionary tracks for stars of these masses only if
overshooting is assumed (i.e.\ the gas that rises with convection moves
a little bit beyond the boundary given by the Schwarzschild
criterion for convection). HR\,6902 thus proves the importance of 
overshooting (Schr\"oder et al.\ 1997). A star of 3\,$\msun$
with the nominal luminosity of the secondary (indicated 
in the Figure with a $\bullet$) has an age of about 260\,Myr, which is
older than the full evolution of a star of 4\,$\msun$. Only if the
luminosity of the secondary is less, is its age compatible
with the age of the primary. E.g.\ a  star of 3\,$\msun$ reaches 
$L=100L_\odot$ (well within the measurement accuracy) after 170\,Myr.
%\nocite{spe97}

Whereas tests on non-interacting stars are more or less straightforward,
it is more difficult to constrain processes like loss of mass
and angular momentum from the binary. An example of such a test is
the one on the binary AS\,Eri. This is an Algol-type binary, i.e.\ a
binary in which a giant transfers mass to a more massive unevolved star.
The giant must originally have been the more massive star, but
has become less massive than its companion by transferring mass
to it or by losing mass which then leaves the binary. Refsdal et
al.\ (1974) have shown that the current component masses and orbital
period of AS\,Eri imply that the binary must have lost a significant 
amount of angular momentum during the mass transfer, presumably
with a significant loss of mass.
%\nocite{rrw74}

In Figure\,\ref{verbfa} we can assign several other areas of binary
evolution where more knowledge is required. Single stars lose mass
in the form of a stellar wind; what is the effect of this in
a binary (in e.g.\ phases a-b and e-f)? Does it indeed, as is 
often assumed, widen the binary? Is the stellar wind stronger when a star 
comes close to its Roche lobe (Tout \&\ Eggleton 1988)?
Is mass transfer between the stars accompanied by loss of mass
from the binary (phases b-d)? What happens exactly when the mass
tranfer is dynamically unstable? Do we indeed get a spiral-in (phases
f-g)?
In general, the answers to questions like these can be sought
on one hand in the study of carefully selected binaries -- like
AS\,Eri -- and on the other hand by the statistical study of large
numbers of binaries, combined with population synthesis. 
%\nocite{te88}

\section{Triple and multiple systems}

\begin{figure}
\centerline{\psfig{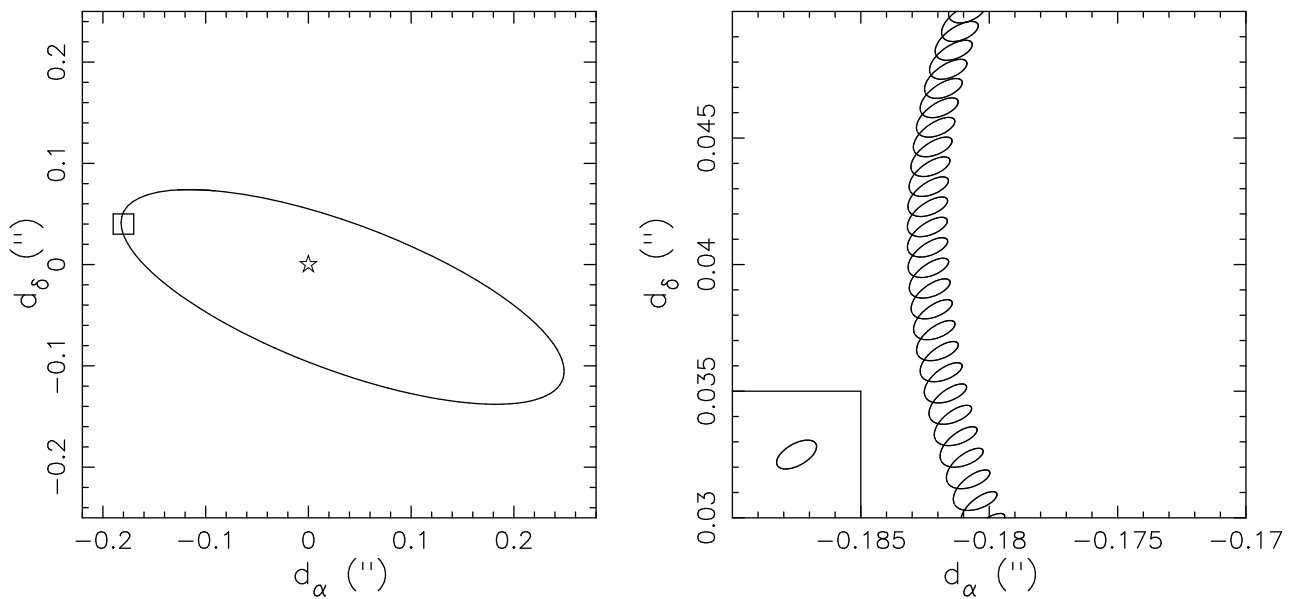}}

\caption[o]{\it Outer and inner orbit of $\kappa$\,Peg
(Muterspaugh et al.\ 2004). The outer orbit, in which the binary
$\kappa$\,Peg\,B revolves in 11.6\,yr around $\kappa$\,Peg\,A is shown left.
Part of the orbit (indicated by a box) is shown enlarged in the right hand
frame, where the motion of the brighter star in $\kappa$\,Peg\,Ba around
(the center of mass with) its companion $\kappa$\,Peg\,Bb becomes visible.
Thus the inner orbit of Ba-Bb can be determined if the outer orbit AB 
is measured with sufficient accuracy. The resulting inner orbit Ba-Bb
is shown as an inset.
     \label{verbfg}}
\end{figure}

Perhaps the first thing to note about triple and multiple systems is
that they are common! Several naked eye stars are very complex
systems. An example is Castor ($\alpha$\,Gem) which consists of three
stars, each of which is a binary (Heintz, 1988): Castor\,A, a binary
of an A star with an unknown companion in a period of 2.9\,d is in an
orbit of 467\,yr with Castor\,B, a binary of an A star with unknown
companion in an orbit of 9.2\,d; this quadruple system is in a
$\sim$Myr orbit around Castor\,C, a binary of two M dwarfs in a
19.8\,hr orbit. It is remarkable that the companions of the A stars in
Castor A and B have not been detected so far.
%\nocite{hei88}

The detailed study of the nearby cluster Praesepe finds a ratio of
single to binary to triple stars as $s:b:t=47:30:3$, i.e.\ of 115
stars, 60 are in a binary and 9 in a triple (Mermilliod \&\ Mayor
1999; see also Mermilliod et al.\ 1994, with a list of triple stars in
several clusters).  A speckle interferometry study of the O stars in
the Trapezium of the Orion nebula finds 1.75 companions per primary,
on average, again indicating a high incidence of multiple systems
(Weigelt et al.\ 1999).  And finally, continued study of
radial-velocity orbits of close binaries in the field indicates that
a sizable fraction of them probably are part of multiple systems
(Mayor \&\ Mazeh 1987).
%\nocite{mm99b}\nocite{mdm94}\nocite{wbp+99}\nocite{mm87}

The stability of triple stars has been the subject of much research.
In general, a three-body system is not stable unless it is hierarchical,
i.e.\ two stars form a binary which is much closer than the distance
to the third star. Thanks to some brilliant mathematics by Mardling
(2001) a simple criterion is now available to determine the stability
of any coplanar triple system (see also Sect.\,4 of Mardling \&\ Aarseth 
2001).
%\nocite{mar01}\nocite{ma01}

An interesting aspect of triple systems has recently been discussed
by Muterspaugh et al.\ (2004; see also Lane \&\ Muterspaugh 2004). 
$\kappa$\,Peg is a triple in which
a 5.97\,d binary ($\kappa$\,Peg\,B) is in a 11.6\,yr orbit with a 
third star ($\kappa$\,Peg\,A). The outer orbit has a projected
size of about 0.4$''$, and is easily resolved with the Palomar 
Testbed Interferometer. The distance between the two stars A and B
is determined to a precision of about 40\,$\mu$as: since the short-period
binary has a projected semimajor axis of almost 1\,mas, the observations
provide a visual orbit for the short-period binary (Fig.\,\ref{verbfg}).
%\nocite{mlb+04}\nocite{lm04}

In general then, it appears possible to resolve a binary with a
size smaller than the resolution of an interferometer, 
provided that there is a sufficiently nearby third star
that is used as a reference. This third star may be in a triple with the
binary; but the method does not require this. 

An example of a target that will be accessible to a future
interferometer with $\sim$mas resolution is the blue straggler S\,1082
in the old open cluster M\,67 (Van den Berg et al.\ 2001). This star
is in fact a triple, in which a 1.07\,d eclipsing binary of stars with
masses 2.7 and 1.7 $M_\odot$ is in an orbit of $\sim$3\,yr with a
third star of $\sim1.7M_\odot$. Both the primary in the inner binary
and the third star are blue stragglers of their own account. With
Eq.\,\ref{veq:a} we see that the outer and inner orbits have semimajor
axes of about 5\,mas and 40\,$\mu$as, respectively.
%\nocite{bovs01}

\section{Binaries with neutron stars or black holes}

Visual orbits would be very useful for the study or X-ray binaries
in which a neutron star or black hole accretes matter from a companion
(for a recent review, see Charles \&\ Coe 2004).
Even in systems in which eclipses or ellipsoidal flux variations
give an indication of the inclination of the orbit, the actual value
of the inclination is often very uncertain. Thus, a visual orbit and
with it a more accurate inclination will help to improve the accuracy
of the mass determination.
%\nocite{cc04}

Consider for example the well-studied X-ray binary Cyg X-1, in which
a black hole accretes mass from a high-mass O star. Extensive optical
studies cannot constrain the inclination more accurately than
a range from 28$^\circ$ to 38$^\circ$, corresponding
to a range of the mass of the black hole of 20 to 10\,$M_\odot$
(Gies \&\ Bolton 1986).
For a total mass $\sim50M_\odot$, orbital period of 5.6\,d
and a distance $\simeq2$\,kpc, the semimajor axis of the orbit is 
about 0.1\,mas.
Similarly, the inclination of the 9\,d orbit of Vela X-1 may range from 
73$^\circ$ to 90$^\circ$ allowing masses of the neutron star between 
1.8 and 2.0\,$M_\odot$ and of its O star companion between 23 and 26
$M_\odot$ (Barziv et al.\ 2001). At a distance of 1.4\,kpc the semimajor
axis is 0.18\,mas.
%\nocite{gb86}\nocite{bkk+01}

As an example of a black hole that accretes from a less massive donor,
we consider the widest system GRS\,1915+105: a (14$\pm$4)$M_\odot$ black hole
accreting from a $\sim 1M_\odot$ star in an orbit of 34\,d (Greiner
et al.\ 2001). At a distance of 10\,kpc the semimajor axis is
0.05\,mas. A relatively nearby low-mass X-ray binary is Sco X-1,
in which a neutron star forms a 19\,hr binary with a low-mass donor.
At a distance of 2.8\,kpc (Bradshaw et al.\ 1999) 
its semimajor axis is about 0.01\,mas. 
%\nocite{gcm01}\nocite{bfg99}

We see that to study X-ray binaries, interferometers will have to
reach accuracies down to fractions of a milliarcsecond. For the high-mass
systems, magnitudes are not a problem ($V=8.9$ for Cyg X-1, and
$V=6.9$ for Vela X-1). The low-mass systems are much less bright in the
optical ($K\simeq13$ for GRS\,1915+105 and $V=12.4$ for Sco X-1).
However, the study of these binaries with interferometry will be 
complicated by their intrinsic strong variability, and by the fact that the
size of the stars is about half of the distance between them.
For the low-mass binaries, both objects (mass donor and accretion disk)
have comparable fluxes in the visual, but in high-mass binaries
the accretion disk is very much fainter than the O or B star donor.

The systems discussed so far have donors that fill their Roche lobe
in relatively short orbital periods. Although less conspicuous in X-rays 
than these systems, the majority of high-mass X-ray binaries
may actually be Be stars that irregularly transfer mass to a neutron
star in much wider orbits (see Fig.\,\ref{verbfa}e). These systems
are being discovered in increasing numbers as transient bright sources
of X-rays (see catalogue by Liu et al.\ 2000). 
With a typical mass of $\sim 15M_\odot$, a period of a year,
and a distance of 2.5\,kpc, a semimajor axis of 1\,mas follows.
The transient nature of these sources implies that the interferometry
must be performed during an outburst, i.e.\ as a target-of-opportunity.
%\nocite{lph01}

Low-mass X-ray binaries may evolve into relatively wide binaries in
which a radio pulsar is accompanied by an undermassive (i.e.\ 0.2-0.4
$M_\odot$) white dwarf. Since the neutron star doesn't emit detectable
optical radiation, the motion of the white dwarf can only be measured
against a third object. Binaries of this type occur in globular
clusters, where third objects are certainly available. An interesting
example is PSR\,1620$-$26, an 11\,ms pulsar in a 191\,d orbit with an
0.3\,$M_\odot$ white dwarf, in the globular cluster M\,4 (Sigurdsson et 
al.\ 2003, and references therein). Its semimajor axis at the cluster
distance of 2.2\,kpc is 0.37\,mas. The problem with this system is its
extreme faintness, at $V=24$.
%\nocite{srh+03}

In conclusion we may say that the study of compact stars in binaries
requires a resolution well below a milliarcsecond. This is true not
only for the binaries with neutron stars and black holes that we discussed
above, but also for the white dwarfs accreting from a low-mass donor
star in cataclysmic variables. A well-studied example is SS\,Cyg, at about
160\,pc (Harrison et al.\ 1999). Its semi-major axis corresponds to 0.06\,mas.
%\nocite{hms+99}

\section{Closing remarks}

A next generation of interferometers, which can resolve binaries down to
the milliarcsecond level, will be able to determine the visual orbits
of enough binaries (several thousand) to determine the properties
(masses, orbital period) with which young binaries are born. 
This requires a very large observing programme (comparable to e.g.\ the
CORAVEL survey), and thus probably a dedicated telescope.
The number of observations may be limited, first by limiting the survey
to binaries that evolve within the Hubble time, i.e.\ with primary masses
higher than 0.8\,$M_\odot$, and that are close enough that the stars
may affect one another's evolution, i.e.\ with orbital periods less than
30-50\,yr. Also, by selecting binaries with known orbital periods and 
orbital phases, one may be able to time the interferometric observations
cleverly for faster results. Such binaries could be found in clusters
of stars, where multi-object spectroscopy would be efficient in
determining large numbers of orbits.

Knowledge about the evolution of binaries may be more difficult to obtain;
on the other hand, by selecting specific binaries one can obtain
interesting results with a relatively small programme.

The study of X-ray binaries requires such small resolution for stars often
very faint that it may not be possible with the next generation of
instruments. In some cases, indirect resolution of an orbit by use of a
third star may be possible.

A question which will have to be adressed is in how far the GAIA
satellite will already adress the questions of binary formation and
evolution.  Its fantastic resolution (down to 3$\mu$as) and very large
number (10$^9$) of observed stars make this an important mission
for binary studies (e.g.\ Zwitter \&\ Munari 2004).  Especially
large-scale programmes for groundbased interferometry should be aimed
to complement, rather than duplicate, the GAIA results.

%
% USE A SECTION WITHOUT NUMBER FOR THE ACKNOWLEDGEMENTS
%
%\section*{Acknowledgements}
%This research was supported in part by contract ARC 90/94-140 ``Action de
%Recherche Concert\'ee de la Communaut\'e Fran\c{c}aise'' (Belgium).
%
% BEGIN THE REFERENCE LIST WITH \beginrefer
% USE \refer BEFORE THE REFERENCES AND BEGIN A NEW PARAGRAPH AFTER THE 
% REFERENCE !
% DO NOT FORGET TO END THE LIST WITH \endrefer
%
 
\beginrefer
\refer
Barziv, O., Kaper, L., van Kerkwijk, M., Telting, J., \& van Paradijs, J. 2001,
  A\&A, 377, 925

\refer
Binney, J. \& Merrifield, M. 1998, Galactic Astronomy (Princeton University
  Press)

\refer
Boden, A., Lane, B., Creech-Eakman, M., et~al. 1999, ApJ, 527, 360

\refer
Bradshaw, C., Fomalont, E., \& Geldzahler, B. 1999, ApJ, 21, L121

\refer
Charles, P. \& Coe, M. 2004, in Compact stellar {X}-ray sources, ed. W.~Lewin
  \& M.~van~der Klis (Cambridge University Press), in press
  (astro-ph/0308020)

\refer
Covino, E., Catalano, S., Frasca, A., et~al. 2000, A\&A, 361, L49

\refer
d{'}Antona, F. \& Mazzitelli, I. 1994, ApJS, 90, 467

\refer
Duquennoy, A. \& Mayor, M. 1991, A\&A, 248, 485

\refer
Gies, D. \& Bolton, C. 1986, ApJ, 304, 371

\refer
Greiner, J., Cuby, J., \& McCaughrean, M. 2001, Nature, 414, 522

\refer
Harrison, T., McNamara, B., Szkody, P.,  et~al. 1999, ApJ, 515, L93

\refer
Hartkopf, W., Mason, B., \& Worley, C. 2001, AJ, 122, 3472

\refer
Heintz, W. 1988, PASP, 100, 834

\refer
Lane, B. \& Muterspaugh, M. 2004, ApJ, 601, 1129

\refer
Liu, Q.-Z., van Paradijs, J., \& van~den Heuvel, E. 2000, A\&AS, 147, 25

\refer
Mardling, R. 2001, in Evolution of Binary and Multiple Star Systems, ed.
  P.~Podsiadlowski \& et~al., ASP Conference Series 229, 101

\refer
Mardling, R. \& Aarseth, S. 2001, MNRAS, 321, 398

\refer
Mason, B., Gies, D., Hartkopf, W., Bagnuolo, W., ten Brummelaar, T., \&
  McAlister, H. 1998, AJ, 115, 821

\refer
Mayor, M. \& Mazeh, T. 1987, A\&A, 171, 157

\refer
Mermilliod, J.-C., Duquennoy, A., \& Mayor, M. 1994, A\&A, 283, 515

\refer
Mermilliod, J.-C.and~Mayor, M. 1999, A\&A, 352, 479

\refer
Muterspaugh, M., Lane, B., Burke, B., Konacki, M., \& Kulkarni, S. 2004, SPIE
  Conference: New Frontiers in stellar interferometry, 5491, in press 
  (astro-ph/0407088)

\refer
Quirrenbach, A. 2001, in The formation of binary stars, IAU Symposium 200, ed.
  H.~Zinnecker \& R.~Mathieu (ASP), 539--546

\refer
Refsdal, S., Roth, M., \& Weigert, A. 1974, A\&A, 36, 113

\refer
Schertl, D., Balega, Y., Preibisch, T., \& Weigelt, G. 2003, A\&A, 402, 267

\refer
Schr{\"o}der, K.-P., Pols, O., \& Eggleton, P. 1997, MNRAS, 285, 696

\refer
Sigurdsson, S., Richer, H., Hansen, B., Stairs, I., \& Thorsett, S. 2003,
  Science, 301, 193

\refer
Steffen, A., Mathieu, R., Lattanzi, M., et~al. 2001, AJ, 122, 997

\refer
Tout, C. \& Eggleton, P. 1988, MNRAS, 231, 823

\refer
van~den Berg, M., Orosz, J., Verbunt, F., \& Stassun, K. 2001, A\&A, 375, 375

\refer
Verbunt, F. 1993, ARA\&A, 31, 93

\refer
Weigelt, G., Balega, Y., Preibisch, T., et~al. 1999, A\&A, 347, L15

\refer
Zwitter, T. \& Munari, U. 2004, Rev. Mex. A.A. 21, 251

\endrefer           
\end{document}